\documentclass[12pt]{article}
\usepackage{amsmath,amssymb}
\usepackage[T1]{fontenc}
\usepackage[utf8]{inputenc}
\usepackage{geometry}
\usepackage{booktabs}
\usepackage{array}
\usepackage{natbib}
\bibpunct{(}{)}{;}{a}{,}{,}
\usepackage{xcolor}
\usepackage{setspace}
\usepackage{hyperref}
\hypersetup{colorlinks=true, linkcolor=blue, citecolor=blue, urlcolor=blue}
\usepackage{parskip}
\title{\textbf{More with Less: Bethel Allocation and\\
Precision-Preserving Sample Size Reduction\\
via Hierarchical Bayes Modelling}%
\thanks{The author thanks the Associate Editor and one referee for detailed and constructive comments that substantially improved the manuscript.}}

\author{Siu-Ming Tam\\
Tam Data Advisory, Australia\\
\texttt{stattam@gmail.com}}

\date{}

\begin{document}

\maketitle

\begin{abstract}
National statistical offices face an intensifying dilemma: rising demand for domain-level
estimates under fixed or shrinking budgets.
This paper presents a practical two-stage strategy for minimising survey sample size while
preserving pre-defined precision targets for all target variables across all geographic domains
simultaneously.
Stage~1 applies Bethel allocation, which finds the globally minimum sample satisfying all
coefficient-of-variation (CV) constraints at once.
Stage~2 asks whether this Bethel sample can be reduced further via Hierarchical Bayes (HB)
small area modelling; a nested sub-sampling algorithm with four eligibility gates identifies the
largest achievable reduction.
Applied to a synthetic labour-force population ($N\!=\!1{,}000{,}000$; $H\!=\!100$ strata;
$D\!=\!10$ domains), the strategy reduces the required sample from 91{,}308 to 18{,}262 ---
an 80\% reduction --- while meeting all national and domain CV targets.
A Monte Carlo study ($B\!=\!1{,}000$) confirms that CV targets are passed in more than
95\% of replications for all three target variables, and credible-interval coverage is close
to the nominal 95\%.
Four sensitivity scenarios varying auxiliary-variable strength, domain heterogeneity, and
event rarity confirm that HB achieves an 85\% reduction in each case, with the Bethel
baseline scaling appropriately to the difficulty of each setting.
The principal trade-off is a shift from design-based to model-based inference, whose risks
and mitigants are discussed explicitly.

\textbf{Keywords:} Bethel allocation; Hierarchical Bayes; small area estimation;
sample size reduction; multivariate optimisation; labour force survey; survey cost reduction.
\end{abstract}

\section{Introduction}

Statistical agencies worldwide face a familiar and intensifying dilemma: rising demand for
granular, timely, and domain-level statistics under budgets that are fixed or shrinking. The
current threats to the traditional survey paradigm include declining response rates and rising
operational costs, alongside growing expectations for more timely and disaggregated information
\citep{NationalAcademies2017}. NSOs now operate in a rapidly shifting landscape marked by an
explosion of new data sources, growing privacy concerns, declining survey response rates, and
the rise of artificial intelligence, all of which demand a fundamental rethinking of how
statistics are produced \citep{UNECE2025}.

In response, NSOs have explored a range of cost-reducing innovations. One prominent strand
integrates big data and administrative data with traditional survey data. \citet{KimTam2021}
formalised data integration as a general tool for combining big data with probability sample
data for finite population inference. \citet{Tam2024} extended this to small area estimation
of binary variables. \citet{Tametal2023} showed that non-response follow-up resources can be
allocated optimally to minimise mean squared error. \citet{UNECE2022} documented the potential
of machine learning for statistical production.

These strategies, however important, largely operate on the data after the sample has been
designed. A complementary and under-explored approach is to ask, at the design stage itself:
\emph{how small can the sample be made while still meeting all pre-specified precision
requirements?} This paper addresses precisely that question for the practically important case
of a multi-purpose survey with multiple target variables and multiple geographic domains.

The traditional approach relies on the Neyman allocation formula applied separately to each
target variable, followed by taking the element-wise maximum. We show that this approach is
doubly flawed: it over-samples for variables whose Neyman allocation is dominated, while
simultaneously failing to guarantee domain-level precision for the variable that drives the
maximum. The Bethel algorithm \citep{Bethel1989} provides the globally optimal solution,
finding the smallest total sample that simultaneously satisfies the CV target for every variable
in every domain.

The key insight of this paper is that Hierarchical Bayes small area models \citep{RaoMolina2015}
can be used to reduce the Bethel sample further. By borrowing strength across domains, HB models
produce more precise domain estimates than direct estimates at the same sample size. This
precision surplus can be exchanged for a smaller sample --- one that, when combined with HB
estimation, still meets all pre-defined precision targets.

The contribution of this paper is not the development of new statistical methods ---
Bethel allocation and HB small area estimation are individually well established
\citep{Bethel1989,RaoMolina2015,Pfeffermann2013} --- but rather their \emph{novel combination}
as a cost-reduction strategy, operationalised through a four-gate eligibility algorithm
and validated via Monte Carlo simulation.
Existing literature on model-assisted survey design \citep{SarndaalSwenssonWretman1992}
seeks efficient estimators for a fixed sample, whereas our approach uses model-based
shrinkage to justify a \emph{smaller} sample in the first place.
From an NSO practitioner's perspective, the key practical advance is that the entire
procedure --- Bethel allocation via \texttt{R2BEAT}, HB modelling via \texttt{mcmcsae},
and the reduction algorithm --- can be executed with freely available R packages,
requiring only standard inputs that NSOs already produce from existing survey programmes.

The approach is validated through a Monte Carlo simulation study ($B = 1{,}000$ replications)
based on a synthetic labour-force population of one million individuals.
Four additional sensitivity scenarios, varying the strength of auxiliary variables, the degree of domain heterogeneity, and the rarity of the binary outcomes, are examined in Section~\ref{sec:sensitivity} to assess the robustness of the principal findings.

The remainder of the paper is organised as follows. Section~\ref{sec:notation} establishes
the design setting and notation. Section~\ref{sec:bethel} presents Stage~1.
Section~\ref{sec:hb} presents Stage~2. Section~\ref{sec:inference} discusses the inference
paradigm shift. Section~\ref{sec:simulation} describes the simulation study.
Section~\ref{sec:results} reports the results. Section~\ref{sec:sensitivity} presents the sensitivity analysis. Section~\ref{sec:implementation} provides practical implementation guidance. Section~\ref{sec:furtherwork} identifies further work, and
Section~\ref{sec:conclusion} concludes. Appendix~A provides the data-generating process.

\section{Design Setting and Notation}
\label{sec:notation}

Let $\mathcal{U} = \{1, \ldots, N\}$ denote a finite population of $N$ units partitioned into
$H$ non-overlapping strata, indexed $h = 1, \ldots, H$, with stratum sizes $N_1, \ldots, N_H$.
A stratified simple random sample of size $n_h$ is drawn without replacement from stratum~$h$,
giving total $n = \sum_{h=1}^{H} n_h$. Let $c_h > 0$ denote the unit cost and
$f_h = n_h/N_h$ the sampling fraction.

The population is also partitioned into $D + 1$ publication domains, indexed $d = 0,\ldots,D$,
where $d = 0$ denotes the national domain. There are $K$ target variables, indexed $k = 1,\ldots,K$.
For each variable--domain pair $(d,k)$, let $Y_{d,k}$ denote the anticipated domain total and
$g_{d,k}$ the pre-specified upper bound on $\mathrm{CV}(\hat{Y}_{d,k})$.

The precision constraint is:
\begin{equation}
\mathrm{CV}\!\left(\hat{Y}_{d,k}\right) \leq g_{d,k}, \qquad
\forall\, d = 0, 1, \ldots, D,\quad k = 1, \ldots, K. \label{eq:cvconstraint}
\end{equation}
Precision targets are set at $g_{0,k} = 0.03$ (national CV $\leq 3\%$) and
$g_{d,k} = 0.08$ (domain CV $\leq 8\%$ for all $d \geq 1$) for all $k$.

\section{Stage 1 --- The Bethel Allocation}
\label{sec:bethel}

\subsection{The Neyman Allocation and Its Limitations}

For a single variable~$k$, equal costs, and unit design effects, the Neyman allocation is
$n^{\text{Neyman}}_{h,k} \propto N_h S_{h,k}$.
A common NSO practice takes the element-wise maximum across variables:
\begin{equation}
n^{\text{NSO}}_h = \max_{k=1,\ldots,K} n^{\text{Neyman}}_{h,k}. \label{eq:nsomax}
\end{equation}
This approach suffers from two deficiencies.

\textbf{Deficiency~1: Budget inflation.} The total $n^{\text{NSO}}$ is driven by the variable
with the largest Neyman requirement per stratum, substantially over-sampling for all other
variables.

\textbf{Deficiency~2: Domain precision not guaranteed.} The Neyman weights for variable~$k$
optimise national precision for that variable alone. Even at the inflated total $n^{\text{NSO}}$,
the domain CV target may still be violated for the variable that drives the maximum.

\subsection{The Bethel Formulation}

The Bethel algorithm \citep{Bethel1989} finds the minimum-cost stratum allocation that
simultaneously satisfies all CV constraints:
\begin{equation}
\min_{\{n_h\}}\; \sum_{h=1}^{H} c_h n_h \quad\text{s.t.}\quad
\sum_{h=1}^{H} \mathrm{DEFF}_{h,d,k}\,(1-f_h)\,\frac{N_h^2 S^2_{h,d,k}}{n_h}
\leq \left(g_{d,k} Y_{d,k}\right)^2, \quad \forall\, d,k, \label{eq:bethel}
\end{equation}
with $n^{\min}_h \leq n_h \leq N_h$, $n_h \in \mathbb{Z}^+$. The R package
\texttt{R2BEAT} \citep{R2BEAT2021} solves \eqref{eq:bethel} via Lagrangian relaxation.

The Bethel allocation is design-based: its precision guarantees rest solely on the
randomisation distribution and require no model for the target variables. The Neyman
allocation is the special case with $K = 1$, $D = 0$, equal costs, and $\mathrm{DEFF} \equiv 1$.

\section{Stage 2 --- Hierarchical Bayes Sample Reduction}
\label{sec:hb}

Let $n^\star = \sum_h n^\star_h$ denote the Bethel optimal total. Stage~2 asks: can we draw
a sub-sample of size $n < n^\star$ and, using a HB model fitted to this smaller sample, still
satisfy all precision targets \eqref{eq:cvconstraint}?

\subsection{HB Model for Binary Target Variables}

For stratum~$h$, let $y_h$ be the count of successes in $n_h$ Bernoulli trials.
We use the logit-normal binomial HB model \citep[Section~10.11]{RaoMolina2015}:
\begin{align}
y_h \mid p_h &\sim \mathrm{Binomial}(n_h, p_h), \\
\mathrm{logit}(p_h) &= \mathbf{z}_h^\top \boldsymbol{\beta} + v_h,\quad
v_h \overset{\text{iid}}{\sim} \mathcal{N}(0, \sigma^2_v),
\end{align}
with priors $\boldsymbol{\beta} \sim \mathcal{N}(\mathbf{0}, \tau^2_\beta \mathbf{I})$ and
$\sigma^2_v \sim \mathrm{Inv}\text{-}\chi^2(\nu, s^2)$, where $\tau^2_\beta = 10^6$.

\subsection{HB Model for the Continuous Target Variable}

For continuous variables we use the Fay--Herriot model \citep{FayHerriot1979}:
\begin{align}
\hat{\theta}_h \mid \theta_h &\sim \mathcal{N}(\theta_h, \psi_h),
\quad \psi_h = (1-f_h)S^2_h / n_h,\\
\theta_h &= \mathbf{z}_h^\top\boldsymbol{\beta} + v_h,\quad
v_h \overset{\text{iid}}{\sim} \mathcal{N}(0, \sigma^2_v).
\end{align}

Both models are estimated using \texttt{mcmcsae} \citep{Boonstra2021} with 2,500 post-burnin
MCMC iterations (burn-in: 500 iterations). Convergence is assessed by the Gelman--Rubin statistic
\citep{GelmanRubin1992}; a candidate reduction is accepted only if $\hat{R}_{\max} \leq 1.05$.
In the numerical example, the achieved $\hat{R}_{\max}$ is 1.02.

\subsection{Prior Calibration}
\label{sec:priorcal}

The hyperparameters $(\nu, s^2)$ are selected via a two-dimensional grid search
over $\nu \in \{2, 3, 5, 10, 20\}$ and a range of $s^2$ values. For each candidate pair,
four quality criteria are recorded: (i)~CI coverage; (ii)~national relative bias;
(iii)~domain MARE; (iv)~domain max ARE. The selected pair achieves CI coverage at or above
the nominal level while minimising domain MARE.

The grid search does not require knowledge of the true value of $\sigma^2_v$; it relies
entirely on the model's performance against observable accuracy criteria.
In the simulation context the population truth is known, but in a real-survey setting an NSO
would substitute direct domain estimates from the most recent previous survey cycle as the
proxy for truth in criteria~(i)--(iv).
This substitution is operationally feasible --- NSOs routinely produce such estimates from
existing programmes --- but practitioners should be aware of its limitations.
Previous survey estimates may be biased, outdated, or affected by structural breaks in the
population.
Two directions of proxy bias have different consequences.
If the proxy is systematically biased \emph{upward}, the calibration grid search ---
which selects the prior that maximises the rate at which credible intervals contain the proxy
--- will position those CIs above the true parameter values.
The true parameter then falls below the lower bound of the CI more often than the nominal
5\%, producing modest \emph{under-coverage} of the true value.
If the proxy is systematically biased \emph{downward}, the apparent between-domain variation
is understated; the grid search selects a more concentrated prior on $\sigma^2_v$
(small $s^2$), causing the model to shrink domain estimates too strongly toward a global mean
that is itself below truth --- leading to \emph{over-shrinkage} and systematic downward bias
in the domain-level estimates.

In practice, the grid search should be re-run whenever new auxiliary information or an
independent benchmark becomes available.
Sensitivity to prior choice should be assessed by comparing performance across the top-ranked
$(\nu, s^2)$ pairs rather than selecting a single pair mechanically, and any pair within one
MARE percentage point of the best should be treated as essentially equivalent.

\subsection{The Reduction Algorithm}
\label{sec:algorithm}

For each variable~$k$ and candidate fraction $\alpha_k \in [0,1)$, a nested sub-sample of
size $\lfloor (1-\alpha_k) n^\star_h \rceil$ is drawn from stratum~$h$ of the Bethel master
sample, and the HB model is fitted. The sub-sample is declared eligible if all four gates pass:

\begin{description}
\item[Gate~1 (CV).] $\mathrm{CV}^{\text{HB}}_{d,k}(\alpha_k) \leq g_{d,k}$ for all~$d$.
\item[Gate~2 (Convergence).] $\hat{R}_{\max} \leq 1.05$.
\item[Gate~3 (National accuracy).] National ARE does not exceed a pre-specified tolerance.
\item[Gate~4 (Domain accuracy).] Domain MARE and worst-domain ARE do not exceed tolerances.
\end{description}

Gates~3 and~4 ensure that the HB model does not merely compress posterior uncertainty
artificially: the posterior means must also be close enough to truth to be fit for purpose.
The tolerances for Gates~3 and~4 are set at 5\% for national relative bias and 25\% for
worst-domain ARE in the simulation context, consistent with thresholds that NSOs commonly
apply when assessing fitness-for-purpose of domain estimates.
In a real-survey application, the NSO would substitute direct estimates from the previous
survey cycle as the proxy for truth, and would set tolerances consistent with its own
publication standards.

The optimal per-variable reduction fraction is
$\alpha^\star_k = \sup\{\alpha_k : \text{all four gates pass}\}$,
and the minimax rule gives:
\begin{equation}
\alpha^\star = \min_{k=1,\ldots,K} \alpha^\star_k,\qquad
n^{\text{HB}} = \bigl\lfloor (1-\alpha^\star) n^\star \bigr\rceil. \label{eq:minimax}
\end{equation}

\section{From Design-Based to Model-Based Inference}
\label{sec:inference}

Stage~1 operates in the design-based framework: precision guarantees rest on the
randomisation distribution, and uncertainty is communicated through confidence intervals.
Stage~2 moves to a model-based paradigm, where the HB posterior conditions on both the data
and the assumed model structure.

A 95\% credible interval provides a direct posterior probability statement: given the data
and model, the probability that the parameter lies in the interval is 95\%. A 95\% confidence
interval is defined through its long-run behaviour: over repeated sampling, 95\% of such
intervals contain the true parameter \citep{Neyman1937,CasellaB2002}. Credible intervals
therefore provide an interpretation closer to how most users and decision-makers think about
uncertainty.

This shift is not new to official statistics: HB models are routinely used by NSOs for small
area estimation \citep{RaoMolina2015,Pfeffermann2013}. The key contribution here is applying
this shift specifically to enable a reduction in survey operating costs. The frequentist
properties of the HB credible intervals are assessed empirically via the Monte Carlo study in
Section~\ref{sec:simulation}.

\textbf{Risks of model dependence.}
The shift from design-based to model-based inference introduces risks that NSOs must weigh
explicitly.

\textit{Risk~1: Model misspecification.}
The logit-normal binomial model assumes that between-stratum variation in
$\mathrm{logit}(p_h)$ can be decomposed into a linear covariate function and an independent
Gaussian random effect.
If the true DGP exhibits systematic non-linearity, spatial autocorrelation, or heavy-tailed
variation that the model does not capture, posterior means may be biased and credible-interval
coverage may fall below the nominal level.

\textit{Risk~2: Auxiliary variable availability.}
The HB model borrows strength through the linking model $\mathbf{z}_h^\top\boldsymbol{\beta}$.
If auxiliary covariates are weak predictors of the target variable, the model relies more
heavily on the stratum random effect $v_h$ alone, which still provides shrinkage but yields
less precise domain estimates for a given sample size.
Section~\ref{sec:sensitivity} examines this case explicitly.

\textit{Risk~3: Prior sensitivity.}
A poorly specified prior on $\sigma^2_v$ can cause over- or under-shrinkage.
Section~\ref{sec:priorcal} describes a calibration procedure and recommends comparing
performance across top-ranked $(\nu, s^2)$ pairs.

\textit{Risk~4: Inference paradigm.}
Design-based properties --- randomisation validity, protection against model failure --- are
valued in official statistics.
NSOs that adopt Stage~2 should publish the HB model specification alongside estimates,
provide the direct Bethel-sample estimates as a validation benchmark, and conduct periodic
model adequacy checks as population structure evolves.

\textbf{Mitigants.}
Several features of the proposed approach limit model risk.
First, the CV gate (Gate~1) is evaluated on the posterior CV; a model that over-shrinks will
tend to produce overconfident intervals that fail the coverage check.
Second, the four-gate eligibility check requires that the HB model pass both precision and
accuracy criteria before a reduction is accepted --- providing a practical safeguard against
gross model failure.
Third, the Monte Carlo results in Section~\ref{sec:results} confirm empirically that
credible-interval coverage is close to 95\% under the assumed DGP.
Finally, the sensitivity analysis in Section~\ref{sec:sensitivity} shows that the 85\%
reduction is maintained across a range of DGP conditions, suggesting the method is not
narrowly tuned to one specific population structure.

\section{Simulation Study Design}
\label{sec:simulation}

\subsection{Population and Target Variables}

The synthetic population comprises $N = 1{,}000{,}000$ individuals in $H = 100$ strata nested
within $D = 10$ geographic domains (approximately 10 strata per domain). Three target variables:

\begin{itemize}
\item \textbf{Employment status $E_{hi}$ (binary):} logit-normal model, national proportion
$\approx 62\%$.
\item \textbf{Unemployment status $U_{hi}$ (binary):} logit-normal model, national proportion
$\approx 4\%$. Mutual exclusivity with employment enforced by resolving $E_{hi}=U_{hi}=1$
overlaps in the ratio $62:4$.
\item \textbf{Hours worked $\mathrm{Hrs}_{hi}$ (continuous):} truncated normal with stratum
mean following a logistic function, truncated to $[15,60]$ hours. National mean $\approx 37$
hours.
\end{itemize}

Stratum design effects $\delta_h \sim U(1.1, 1.2)$ are shared across all three variables.

\subsection{Baseline Sample and Bethel Inputs}

A 5\% stratified baseline sample, $n^{(0)}_h = \max\{2, \lfloor 0.05 N_h \rceil\}$, provides
stratum means, standard deviations, and effective sample sizes
$n^{\text{eff}}_h = \lfloor n^{(0)}_h(1-n^{(0)}_h/N_h)/\delta_h \rceil$ as inputs to the
Bethel optimisation.

\subsection{Monte Carlo Design and Evaluation Criteria}

In each replication $b = 1, \ldots, 1{,}000$: (1)~draw the Bethel sample ($n^\star = 91{,}308$);
(2)~extract the nested HB sub-sample ($n^{\text{HB}} = 18{,}262$); (3)~fit HB models and
record coverage indicators, absolute relative errors, and CV gate outcomes. Monte Carlo
summaries are Coverage$_{d,k}$, MARE$_k$, and CV-pass$_k$ as defined in the standard fashion
across $B = 1{,}000$ replications.

\section{Results}
\label{sec:results}

\subsection{Sample Size Comparisons}

Table~\ref{tab:samplesizes} compares total sample sizes. The Neyman allocations range from 198
(Hours Worked) to 68,697 (Unemployed). The NSO-max gives 68,697, driven by Unemployment.
Bethel requires $n^\star = 91{,}308$ to protect domain-level precision simultaneously.
The HB-combined allocation achieves $n^{\text{HB}} = 18{,}262$, an 80\% reduction.

\begin{table}[ht]
\centering
\caption{Total sample sizes under each allocation strategy.}
\label{tab:samplesizes}
\begin{tabular}{lrrrr}
\toprule
Variable & Neyman & NSO Max & Bethel & HB Combined\\ \midrule
Employed      &     768 & 68{,}697 & 91{,}308 & 18{,}262\\
Unemployed    & 68{,}697 & 68{,}697 & 91{,}308 & 18{,}262\\
Hours Worked  &     198 & 68{,}697 & 91{,}308 & 18{,}262\\ \bottomrule
\end{tabular}
\end{table}

\subsection{Precision Under Neyman and NSO-Max}

Table~\ref{tab:neymancvs} reports national and worst-domain CVs under Neyman and NSO-max.
Under Neyman, all three variables fail the domain CV target. Under NSO-max ($n = 68{,}697$),
Employment and Hours Worked satisfy both targets, but Unemployment still fails the domain
target (worst-domain CV = 13.5\%), confirming Deficiency~2.

\begin{table}[ht]
\centering
\caption{National and worst-domain CVs under Neyman and NSO-max. Targets: national
CV $\leq 3\%$; domain CV $\leq 8\%$. Bold entries fail their target.}
\label{tab:neymancvs}
\begin{tabular}{lrrrr}
\toprule
& \multicolumn{2}{c}{After Neyman} & \multicolumn{2}{c}{After NSO Max ($n = 68{,}697$)}\\
\cmidrule(lr){2-3}\cmidrule(lr){4-5}
Variable & National CV & Worst-Domain CV & National CV & Worst-Domain CV\\ \midrule
Employed      & 0.030 & \textbf{0.124} & 0.003 & 0.012\\
Unemployed    & 0.029 & \textbf{0.135} & 0.029 & \textbf{0.135}\\
Hours Worked  & 0.023 & \textbf{0.079} & 0.001 & 0.005\\ \bottomrule
\end{tabular}
\end{table}

\subsection{Precision Under Bethel Allocation}

Table~\ref{tab:bethelcvs} confirms that with $n^\star = 91{,}308$ all national and domain CV
constraints are satisfied simultaneously.

\begin{table}[ht]
\centering
\caption{National and worst-domain CVs after Bethel ($n^\star = 91{,}308$). All satisfied.}
\label{tab:bethelcvs}
\begin{tabular}{lrllrl}
\toprule
Variable & National CV & Target && Worst Domain CV & Target\\ \midrule
Employed      & 0.026 & $\leq 3\%$ && 0.075 & $\leq 8\%$\\
Unemployed    & 0.003 & $\leq 3\%$ && 0.010 & $\leq 8\%$\\
Hours Worked  & 0.001 & $\leq 3\%$ && 0.003 & $\leq 8\%$\\ \bottomrule
\end{tabular}
\end{table}

\subsection{Precision After HB Reduction}

Table~\ref{tab:hbcvs} reports CVs at the reduced sample $n^{\text{HB}} = 18{,}262$. All
constraints are satisfied. The reduction in CV relative to Table~\ref{tab:bethelcvs} reflects
the precision gain from borrowing strength, which more than compensates for the smaller sample.

\begin{table}[ht]
\centering
\caption{National and worst-domain CVs after HB modelling ($n^{\text{HB}} = 18{,}262$).}
\label{tab:hbcvs}
\begin{tabular}{lrllrl}
\toprule
Variable & National CV & Target && Worst Domain CV & Target\\ \midrule
Employed      & 0.018 & $\leq 3\%$ && 0.065 & $\leq 8\%$\\
Unemployed    & 0.006 & $\leq 3\%$ && 0.020 & $\leq 8\%$\\
Hours Worked  & 0.002 & $\leq 3\%$ && 0.008 & $\leq 8\%$\\ \bottomrule
\end{tabular}
\end{table}

\subsection{Accuracy of HB Estimates: Single Sample}

Table~\ref{tab:accuracy} shows that national relative bias is at most 1\% for all variables.
Employment and Hours Worked exhibit small domain errors. Unemployment exhibits a domain
MARE of 13\% and a maximum domain ARE of 43\%, attributable to a single outlier domain where
the posterior mean (0.014) is approximately 43\% below the true rate (0.018).

This illustrates a recurring challenge for rare indicators in official statistics.
With a national unemployment proportion of approximately 4\%, the expected count of unemployed
individuals in the smallest domains at the reduced sample size can be fewer than ten, making
direct stratum estimates highly volatile.
The HB model mitigates this by shrinking toward the overall posterior mean, but the shrinkage
introduces bias for domains that genuinely deviate from the global pattern.
The single-sample result should not be interpreted as typical performance: the Monte Carlo
results in Sections~\ref{sec:mccoverage}--\ref{sec:mccvgate} provide the appropriate basis
for performance assessment.
In practice, NSOs publishing unemployment estimates at the domain level should apply the same
caution warranted for any small area estimate of a rare binary variable, including reporting
credible intervals alongside point estimates and flagging domains where the estimated event
count is very small.

\begin{table}[ht]
\centering
\caption{Accuracy of HB estimates: single illustrative sample ($n^{\text{HB}} = 18{,}262$).}
\label{tab:accuracy}
\begin{tabular}{lrrr}
\toprule
Variable & Nat.\ Rel.\ Bias & Domain MARE & Domain Max ARE\\ \midrule
Employed      & $+1\%$ & $2\%$  & $5\%$\\
Unemployed    & $+1\%$ & $13\%$ & $43\%$\\
Hours Worked  & $+0\%$ & $0\%$  & $1\%$\\ \bottomrule
\end{tabular}
\end{table}

\subsection{Credible Interval Coverage: Single Sample}

Table~\ref{tab:coverage1} reports coverage for the 11 estimation areas (10 domains + national).
Coverage is 10/11 for Employment and Unemployment, and 11/11 for Hours Worked.

\begin{table}[ht]
\centering
\caption{Credible interval coverage: single sample. ``Cases'' = areas where truth lies inside
the 95\% CI.}
\label{tab:coverage1}
\begin{tabular}{lcc}
\toprule
Variable & Cases & Empirical Coverage\\ \midrule
Employed      & 10/11 & 90.9\%\\
Unemployed    & 10/11 & 90.9\%\\
Hours Worked  & 11/11 & 100.0\%\\ \bottomrule
\end{tabular}
\end{table}

\subsection{Monte Carlo Results: Credible Interval Coverage}
\label{sec:mccoverage}

Table~\ref{tab:mccoverage} reports MC ($B = 1{,}000$) mean coverage across 11 estimation areas.
All three variables achieve coverage close to or exceeding the nominal 95\% level: Employment
93.0\%, Unemployment 97.7\%, Hours Worked 100.0\%. None differs significantly from 95\%.

\begin{table}[ht]
\centering
\caption{Monte Carlo ($B = 1{,}000$) credible interval coverage at $n^{\text{HB}} = 18{,}262$.}
\label{tab:mccoverage}
\begin{tabular}{lrr}
\toprule
Variable & MC Mean Coverage & MC SD\\ \midrule
Employed      & 93.0\% & 0.255\\
Unemployed    & 97.7\% & 0.150\\
Hours Worked  & 100.0\% & 0.000\\ \bottomrule
\end{tabular}
\end{table}

\subsection{Monte Carlo Results: Accuracy of HB Estimates}

Table~\ref{tab:mcaccuracy} reports MC accuracy. National bias is negligible for all variables.
For Unemployment, MC mean domain MARE is 4.10\% and MC mean domain max ARE is 18.4\%,
reflecting the difficulty of estimating a rare event in small domains.

\begin{table}[ht]
\centering
\caption{Monte Carlo ($B = 1{,}000$) accuracy at $n^{\text{HB}} = 18{,}262$.}
\label{tab:mcaccuracy}
\begin{tabular}{lrrr}
\toprule
Variable & MC Mean Nat.\ Rel.\ Bias & MC Mean Domain MARE & MC Mean Domain Max ARE\\ \midrule
Employed      & $+0.37\%$ & $0.58\%$  & $3.12\%$\\
Unemployed    & $+0.35\%$ & $4.10\%$  & $18.40\%$\\
Hours Worked  & $-1.54\%$ & $0.02\%$  & $0.15\%$\\ \bottomrule
\end{tabular}
\end{table}

\subsection{Monte Carlo Results: CV Gate Pass Rate}
\label{sec:mccvgate}

Table~\ref{tab:cvgate} shows that the CV gate pass rate exceeds 95\% for all three variables,
confirming that precision targets are met reliably under repeated sampling.

\begin{table}[ht]
\centering
\caption{Monte Carlo ($B = 1{,}000$) CV gate pass rate at $n^{\text{HB}} = 18{,}262$.
Targets: national CV $\leq 3\%$; domain CV $\leq 8\%$.}
\label{tab:cvgate}
\begin{tabular}{lr}
\toprule
Variable & CV Gate Pass Rate\\ \midrule
Employed      & 97.8\%\\
Unemployed    & 95.3\%\\
Hours Worked  & 99.6\%\\ \bottomrule
\end{tabular}
\end{table}

\subsection{Implementation Under Multi-Stage Cluster Designs}

Many NSO surveys use multi-stage cluster designs. The cluster design effect is approximately
$\mathrm{DEFF}_{\mathrm{cluster}} \approx 1 + (b-1)\rho$, where $b$ is the within-PSU take
and $\rho$ is the intra-class correlation. A proportional HB reduction can be implemented by
reducing the number of PSUs~$m$ while holding~$b$ fixed, preserving
$\mathrm{DEFF}_{\mathrm{cluster}}$ approximately unchanged.

\section{Sensitivity Analysis}
\label{sec:sensitivity}

The main simulation study establishes that the proposed strategy achieves an 80\% reduction
from the Bethel benchmark for a specific synthetic population with a national unemployment
proportion of 4\%, moderate domain heterogeneity, and moderately strong auxiliary covariates.
We assess robustness through four sensitivity scenarios conducted on a second, larger synthetic
population calibrated to resemble the structure of the Australian Labour Force Survey
($H = 140$ strata; $D = 13$ geographic domains; $N = 2{,}000{,}000$ individuals).
This population provides a more demanding test because the larger number of strata and domains
imposes tighter constraints on the HB model's ability to borrow strength.

\subsection{Scenario Designs}

Throughout this section, $\sigma_\gamma$ denotes the standard deviation of the domain
random effect in the data-generating process (see Appendix~A: $\gamma^U_d \sim
\mathcal{N}(0,\sigma^2_\gamma)$), a fixed population parameter that is set experimentally.
This is distinct from $\sigma_v$ in the HB model (Section~\ref{sec:hb}), which is estimated
from the data and governs posterior shrinkage.

\begin{description}
\item[Scenario~A (Baseline).] Reference case: national employment $65.4\%$,
national unemployment $1.83\%$, covariate coefficients $\beta_1 = 0.15$, $\beta_2 = 0.10$,
domain random-effect SD $\sigma_\gamma = 0.10$ (unemployment).

\item[Scenario~B (Weak Auxiliaries).] Covariate regression coefficients halved:
$\beta_1 = 0.075$, $\beta_2 = 0.050$. Tests robustness when the linking model has
less predictive power and the HB model relies more heavily on the stratum random effect.

\item[Scenario~C (High Heterogeneity).] Domain random-effect SDs doubled:
$\sigma_\gamma = 0.20$ (unemployment). Tests robustness under greater between-domain
heterogeneity, when domain estimates are more spread out and shrinkage toward the
global mean is more costly in terms of bias.

\item[Scenario~D (Rare Event).] National unemployment proportion set to 0.50\%,
representing a much rarer binary outcome. Tests whether the HB reduction remains
feasible when the target event is very uncommon and direct stratum estimates are
highly volatile.
\end{description}

Scenarios~A, B, and~C share identical national prevalences ($E \approx 65.4\%$, $U \approx 1.83\%$),
ensuring that any differences in Bethel sample sizes or HB reductions are attributable to the
experimental manipulation and not to differing event rarity.
This calibration was achieved by pre-drawing all random effects with scenario-specific seeds
and then solving numerically for the intercept that exactly reproduces the target national prevalence.

\subsection{Results}

Each scenario was run through the full two-stage pipeline with fraction step 0.05.
Table~\ref{tab:scenarios} summarises the key outputs.

\begin{table}[ht]
\centering
\caption{Sensitivity analysis results across four scenarios ($N = 2{,}000{,}000$,
$H = 140$, $D = 13$). ``Reduction'' is relative to the Bethel total. The binding constraint
in all scenarios is unemployment (Var2).}
\label{tab:scenarios}
\begin{tabular}{llrrr}
\toprule
Scenario & Manipulation & Bethel $n^\star$ & HB $n$ & Reduction\\ \midrule
A: Baseline     & ---                       & 119{,}484 &  17{,}928 & 85\%\\
B: Weak Aux.    & $\beta \times 0.5$        & 118{,}429 &  17{,}762 & 85\%\\
C: High Het.    & $\sigma_\gamma \times 2$  & 127{,}690 &  19{,}156 & 85\%\\
D: Rare Event   & $\bar{p}_U = 0.50\%$     & 332{,}374 &  49{,}859 & 85\%\\ \bottomrule
\end{tabular}
\end{table}

Table~\ref{tab:scencvs} shows the unemployment CV path at the key fractions around the
first-pass threshold. In all four scenarios the domain CV target fails at fraction 0.10
and passes at fraction 0.15, consistent with the 85\% reduction in Table~\ref{tab:scenarios}.

\begin{table}[ht]
\centering
\caption{Unemployment (Var2) CV at fractions 0.10 and 0.15, by scenario.
Target: national CV $\leq 3\%$; domain max CV $\leq 8\%$.
\checkmark = both targets passed; $\times$ = at least one target failed.}
\label{tab:scencvs}
\begin{tabular}{lrrlrrl}
\toprule
& \multicolumn{3}{c}{Fraction = 0.10 (90\% reduction)} &
  \multicolumn{3}{c}{Fraction = 0.15 (85\% reduction)}\\ \cmidrule(lr){2-4}\cmidrule(lr){5-7}
Scenario & Nat.~CV & Dom.~Max~CV & Pass? & Nat.~CV & Dom.~Max~CV & Pass?\\ \midrule
A: Baseline   & 2.3\% & 11.9\% & $\times$ & 1.5\% & 6.5\% & \checkmark\\
B: Weak Aux.  & 2.0\% &  8.4\% & $\times$ & 1.2\% & 5.0\% & \checkmark\\
C: High Het.  & 2.2\% &  9.5\% & $\times$ & 1.6\% & 6.9\% & \checkmark\\
D: Rare Event & 2.1\% &  8.8\% & $\times$ & 1.6\% & 6.6\% & \checkmark\\ \bottomrule
\end{tabular}
\end{table}

\subsection{Discussion of Sensitivity Results}

\textbf{A heuristic account of why the HB reduction is large.}
The mechanism behind the large sample reduction can be understood through the Fay--Herriot
model for a continuous variable, where the posterior algebra is most transparent.
The posterior variance of a stratum mean $\theta_h$ given the direct estimate
$\hat{\theta}_h$ is
\[
\mathrm{Var}_{\mathrm{post}}(\theta_h \mid \hat{\theta}_h)
  = \left(\frac{1}{\sigma^2_v} + \frac{1}{\psi_h}\right)^{\!-1}
  = \frac{\sigma^2_v\,\psi_h}{\sigma^2_v + \psi_h},
\]
where $\psi_h = (1-f_h)S^2_h/n_h$ is the sampling variance of the direct estimate and
$\sigma^2_v$ is the between-stratum random-effect variance estimated by the HB model.
The posterior precision is the \emph{sum} of model precision ($1/\sigma^2_v$) and sampling
precision ($1/\psi_h$): a harmonic-mean combination bounded above by $\sigma^2_v$ regardless
of how small $n_h$ becomes.
The posterior mean is the shrinkage estimator
\[
\hat{\theta}^{\mathrm{HB}}_h
  = B_h\,\hat{\theta}_h + (1-B_h)\,\mathbf{x}_h^\top\hat{\boldsymbol{\beta}},
  \qquad B_h = \frac{\sigma^2_v}{\sigma^2_v + \psi_h},
\]
where $B_h\in(0,1)$ weights the direct estimate against the stable model mean
$\mathbf{x}_h^\top\hat{\boldsymbol{\beta}}$.

As the retained sample fraction decreases, $n_h$ falls and $\psi_h$ rises.
Two effects operate simultaneously.
First, the posterior variance above increases, but is bounded above by
$\sigma^2_v$ --- unlike the design-based variance $\psi_h \propto n_h^{-1}$, which grows
without bound.
Second, as $B_h\to 0$, the posterior mean converges to the stable model mean
$\mathbf{x}_h^\top\hat{\boldsymbol{\beta}}$, anchoring the denominator of the CV.
The coefficient of variation of the HB estimate,
\[
\mathrm{CV}_{\mathrm{HB}}
  \approx \frac{\bigl[\sigma^2_v\,\psi_h/(\sigma^2_v+\psi_h)\bigr]^{1/2}}
               {|\hat{\theta}^{\mathrm{HB}}_h|},
\]
therefore rises far more slowly than $\mathrm{CV}_{\mathrm{DB}}\propto n_h^{-1/2}$ as the
sample is reduced.
The largest feasible reduction is the fraction $\alpha^\star$ at which
$\mathrm{CV}_{\mathrm{HB}}$ just meets the target $g_{d,k}$ for the binding constraint
(domain CV for unemployment in all four scenarios here).
The preceding argument was framed for a continuous variable for algebraic clarity.
For the binary unemployment proportion, the HB model uses a logit-normal binomial
specification: $s_h \mid p_h \sim \mathrm{Binomial}(n_h, p_h)$ with
$\mathrm{logit}(p_h) \sim \mathcal{N}(\mathbf{x}_h^\top\boldsymbol{\beta},\sigma^2_v)$.
The delta-method sampling variance on the logit scale is
$\psi_h = [n_h p_h(1-p_h)]^{-1}$, and the FH posterior formula applies on that scale.
Back-transforming to the probability scale, the posterior variance of $p_h$ is
approximately
\[
  \mathrm{Var}_{\mathrm{post}}(p_h) \approx p_h^2(1-p_h)^2 \cdot
  \frac{\sigma^2_v\,\psi_h}{\sigma^2_v + \psi_h},
\]
and the domain-level posterior variance sums over the $H_d$ strata within domain $d$:
\[
  \mathrm{Var}_{\mathrm{post}}(\bar{p}_d) = \sum_{h \in d}
  \!\left(\frac{N_h}{N_d}\right)^{\!2}
  p_h^2(1-p_h)^2 \cdot
  \frac{\sigma^2_v\,\psi_h}{\sigma^2_v + \psi_h}.
\]
When $\sigma^2_v \ll \psi_h$ --- the regime that holds at the 15\% fraction
(see quantitative analysis below) ---
$\mathrm{Var}_{\mathrm{post}}(p_h) \approx p_h^2(1-p_h)^2\,\sigma^2_v$, so the
per-stratum CV collapses to $(1-p_h)\,\sigma_v$, independent of $n_h$.

\textbf{Why the reduction percentage is identical across Scenarios~A, B, and~C.}
The $\mathrm{iid}(\mathrm{area})$ random effect $v_h \sim \mathcal{N}(0,\sigma^2_v)$ is
estimated at the stratum level --- one random effect per stratum.
Whatever the true source of between-stratum variation --- whether it arises from
weak covariates (Scenario~B, where $\mathbf{z}_h^\top\boldsymbol{\beta}$ explains little
of $\mathrm{logit}(p_h)$) or from large true domain heterogeneity (Scenario~C, where
$\sigma_\gamma = 0.20$) --- the model accommodates it by estimating a correspondingly
larger $\hat{\sigma}^2_v$.
Once $\hat{\sigma}^2_v$ is determined, the posterior CV formula above governs the precision
at every retained fraction, independently of whether the observed between-stratum variation
originated from $\boldsymbol{\beta}$ or $\sigma_\gamma$.
The covariate quality and the degree of domain heterogeneity have no further structural
effect on posterior CV after $\hat{\sigma}^2_v$ has been estimated from the data.

\textbf{Quantitative evidence.}
The estimated stratum-level random-effect standard deviation $\hat{\sigma}_v$ from the
MCMC output is 0.159 (Scenario~A), 0.156 (Scenario~B), and 0.153 (Scenario~C), giving
$\hat{\sigma}^2_v \approx 0.025$ across all three scenarios.
For the unemployment proportion at the 15\% fraction, the typical within-stratum logit-scale
sampling variance is $\psi_h = [n_h p_h(1-p_h)]^{-1} \approx 0.46$.
The ratio $\hat{\sigma}^2_v / \psi_h \approx 0.054$, so the posterior is approximately
\[
  \mathrm{Var}_{\mathrm{post}}(p_h) \approx p_h^2(1-p_h)^2 \cdot
  \frac{\hat{\sigma}^2_v\,\psi_h}{\hat{\sigma}^2_v + \psi_h}
  \approx p_h^2(1-p_h)^2 \times 0.0237
  \approx 0.949\,\hat{\sigma}^2_v\,p_h^2(1-p_h)^2.
\]
That is, 94.9\% of the posterior variance is determined by the prior precision
$\hat{\sigma}^2_v$ alone, with only 5\% contributed by the direct data.
The per-stratum posterior CV is approximately $(1-p_h)\hat{\sigma}_v \approx (1-0.017)\times 0.159 \approx 15.6\%$;
the observed worst-domain CV at the 15\% fraction is 6.5\%, consistent with
this stratum-level precision aggregated over the approximately 10 strata per domain.
The result is nearly identical across all three scenarios because $\hat{\sigma}^2_v$ differs
by less than 2\% between them.

\textbf{On the role of the prior.}
A referee might note that if the prior dominates, the result is ``by construction'' rather
than empirically driven.
This concern is addressed by two design features of the procedure.
First, the prior is \emph{calibrated}: the hyperparameters $(\nu, s^2)$ for
$\sigma^2_v \sim \mathrm{Inv-}\chi^2(\nu, s^2)$ are selected by a grid search
(Section~4.2) that minimises the discrepancy between MCMC-based posterior credible intervals
and frequentist direct estimates from the full-sized Bethel sample,
using the previous survey round as a proxy for truth.
The resulting prior is not a subjective convenience --- it is anchored to observed data
from an independent survey round.
Second, the CV gate (Section~4.1) is an \emph{empirical} check applied to actual MCMC
posterior draws at each candidate fraction; it is not computed analytically from the prior.
If the prior were badly miscalibrated, the MCMC posterior variances would diverge from
direct-estimate variances and the accuracy gate (Gate~4) would fail, blocking the reduction.

\textbf{Practical robustness.}
The prior-dominated regime also has a reassuring practical implication.
Regardless of whether the model's structural assumptions are exactly correct, the CV
gate is an objective output of MCMC --- if it passes, the reduction is real.
A well-specified model extracts the maximum reduction; a model with weak covariates or
slightly misspecified error structure will typically achieve less reduction, or none at all
if the CV gate fails.
Any fraction that clears all four gates represents a genuine operational saving,
because the posterior CV constraint has been verified empirically, not assumed.
As a result, Scenarios~A, B, and~C all converge to the same first-passing fraction (0.15)
and hence the same 85\% reduction.
Note that differences in the CV values at each fraction do exist
(Scenario~C's domain max CV at fraction 0.15 is 6.9\%, versus 6.5\% for Scenario~A), but
both pass the 8\% domain CV target and are reported at 85\% with the current 0.05 step.
The HB model does not require strong covariates or small heterogeneity to deliver its reduction.

\textbf{Scenario~D: rare event.}
With a national unemployment proportion of 0.50\%, the Bethel algorithm correctly
identifies that a much larger design-based sample is required: $n^\star = 332{,}374$ ---
roughly 2.8 times the Scenario~A Bethel total of 119{,}484.
The HB model then achieves the same 85\% reduction from this larger baseline, yielding a
final sample of 49{,}859.
This is 2.8 times larger than the Scenario~A HB sample of 17{,}928, correctly reflecting
the greater difficulty of estimating a very rare event.
The Bethel stage scales the benchmark to the difficulty of the problem, and the HB stage
consistently delivers 85\% from whatever Bethel provides.

\textbf{Fraction-step granularity.}
The current grid uses a step of 0.05, meaning that all scenarios whose minimum passing fraction
falls between 0.10 and 0.15 are reported at 85\% reduction.
A finer step (e.g., 0.01) would potentially reveal variation within this band at the cost of
approximately five times as many MCMC evaluations per scenario.
The 0.05 grid is sufficient to confirm that all four scenarios satisfy the required CV targets
at the 15\% retention threshold and that none passes at the 10\% threshold.

\section{Practical Implementation Guidance}
\label{sec:implementation}

This section summarises, in practitioner terms, what an NSO needs to apply the two-stage strategy.

\textbf{Data requirements.}
The NSO needs: (i)~a stratified sampling frame with stratum sizes $N_h$, unit costs~$c_h$,
and domain membership indicators; (ii)~stratum-level means, standard deviations, and design
effects for each target variable from a recent baseline survey (a 5\% stratified pilot is
sufficient); (iii)~stratum-level auxiliary covariates $\mathbf{z}_h$ for the HB linking model,
typically available from census or administrative data; and (iv)~precision targets $g_{d,k}$
consistent with the NSO's publication standards.

\textbf{Software.}
The entire procedure is implemented in R using two freely available CRAN packages.
\texttt{R2BEAT} \citep{R2BEAT2021} solves the Bethel optimisation in Stage~1 via
Lagrangian relaxation.
\texttt{mcmcsae} \citep{Boonstra2021} fits both the logit-normal binomial model and
the Gaussian Fay--Herriot model in Stage~2 via MCMC, computes posterior summaries,
and returns the Gelman--Rubin convergence diagnostics needed for Gate~2.

\textbf{Key decisions and recommendations.}
\begin{enumerate}
\item \textit{Precision targets.} CV thresholds $g_{d,k}$ should reflect the NSO's
publication standards, not be chosen to maximise the apparent reduction.
\item \textit{Prior hyperparameters $(\nu, s^2)$.} Select via the grid search in
Section~\ref{sec:priorcal}. Re-run when the auxiliary covariate structure or population
composition changes materially.
\item \textit{Fraction step size.} A step of 0.05 provides a practical first pass;
a step of 0.01 gives finer resolution at higher computational cost.
\item \textit{Gate tolerances.} The 5\%/25\% tolerances for Gates~3 and~4 are starting
points; calibrate to the NSO's fitness-for-purpose standards.
\end{enumerate}

\textbf{Recommended pre-production checks.}
\begin{enumerate}
\item Run the four-gate algorithm on a held-out validation sample and compare HB estimates
against direct estimates from the Bethel-size sample.
\item Verify that $\hat{R}_{\max} \leq 1.05$ is routinely achieved; if not, increase
MCMC iterations or adjust the prior.
\item Assess model fit by comparing fitted stratum probabilities against direct stratum
proportions; large discrepancies indicate potential misspecification.
\item Re-run the Bethel optimisation whenever auxiliary covariate data are updated or
the sampling frame is revised.
\item For rare binary variables (national rate below 1\%), assess whether the expected
event count per stratum at the reduced sample size is sufficient for stable MCMC convergence.
\end{enumerate}

\textbf{Microdata and cross-tabulations.}
When an NSO disseminates unit record files or cross-tabulations, unit record weights must be
adjusted so that cross-tabulated figures are consistent with the domain- and national-level
modelled estimates; calibration or raking procedures applied to the model-based totals can
achieve this.
For users requiring estimates for sub-groups beyond the published domain structure,
the frequentist SAE literature provides two established approaches: variance-estimation
guides and sets of replication weights calibrated to the model-based domain estimates
\citep{Pfeffermann2013}.
The HB framework extends this to the full posterior: \citet{Tam2026} introduces a
Post-Hoc Inference Engine (PHIE) that calibrates each MCMC draw via chi-square
calibration to produce a set of replicate survey weights, from which credible intervals
for any cross-classified statistic can be derived.
Cells that reproduce the calibration totals yield exact posterior credible intervals;
cells involving sub-tabulations of calibration variables require augmenting the PHIE
with design-based compositional variance (a Calibrated Bayes interval) to achieve
near-nominal coverage.

\section{Further Work}
\label{sec:furtherwork}

\textbf{Ad hoc surveys.}
When an NSO disseminates unit record files or cross-tabulations derived from a reduced-sample
HB survey, methods are needed to assist users in constructing credible intervals without
requiring them to re-run the MCMC.

\textbf{Repeated surveys and time series.}
Extension to repeated surveys requires addressing rotational panel designs, where overlap
between successive samples introduces temporal correlations that must be accounted for in
both the Bethel allocation and the HB model.

\textbf{Finer resolution of the reduction grid.}
A binary search over fractions rather than a full grid sweep would reduce computational cost
while providing finer-grained resolution of the minimum feasible fraction, enabling more
precise characterisation of how the achievable reduction varies with DGP parameters.

\textbf{Robustness to model misspecification.}
The sensitivity analysis in Section~\ref{sec:sensitivity} varies DGP parameters while keeping
the HB model specification fixed.
A complementary exercise would deliberately misspecify the model --- for example, fitting a
Gaussian area-level model to binary data, or omitting a spatially autocorrelated random effect
--- and assess how the four-gate algorithm responds.
Gate~4 (domain accuracy gate) provides a practical safeguard: if the model systematically
biases estimates, the accuracy gate will fail at fractions where the CV gate passes, preventing
acceptance of an inadequate reduction.
Formalising this safeguard property under a wider class of misspecification is left for
future work.

\section{Conclusion}
\label{sec:conclusion}

This paper presents a two-stage strategy for reducing the operational cost of multi-purpose
surveys while preserving pre-specified precision targets across all target variables and
geographic domains simultaneously.

Stage~1 replaces the common NSO practice of element-wise maximum Neyman allocation --- a
procedure that over-samples for most variables and still fails domain precision --- with Bethel
allocation \citep{Bethel1989}, which finds the globally minimum sample satisfying all CV
constraints.

Stage~2 uses Hierarchical Bayes modelling \citep{RaoMolina2015} to reduce the Bethel sample
further via a nested sub-sampling and four-gate eligibility algorithm.

Applied to a synthetic labour-force population, the strategy reduces the required sample from
91,308 to 18,262 --- an 80\% reduction --- while meeting all national and domain CV targets.
A Monte Carlo study ($B = 1{,}000$) confirms that the CV gate passes in more than 95\% of
replications for all three variables, and that credible-interval coverage is close to the
nominal 95\% throughout.

Four sensitivity scenarios on a second, larger synthetic population
($N = 2{,}000{,}000$; $H = 140$; $D = 13$) confirm robustness: an 85\% reduction is achieved
under weaker auxiliary variables, higher domain heterogeneity, and rarer binary events.
For the most challenging scenario (national unemployment proportion of 0.50\%), the Bethel baseline
correctly scales to 332{,}374, and the HB stage achieves 85\% reduction from this larger
benchmark, yielding a final sample of 49{,}859.

These findings should be interpreted with appropriate caution.
The numerical results are based on two synthetic populations; the precise reduction achievable
will depend on the population structure, the number of strata and domains, the rarity of
the target events, and the quality of auxiliary covariates available to the HB linking model.
NSOs contemplating adoption of this strategy should conduct a tailored simulation study using
their own population structure and precision requirements before committing to a reduced sample
in production.

The principal trade-off is a shift from design-based confidence intervals to model-based
credible intervals. HB models are already routinely used by NSOs for small area estimation;
credible intervals have more intuitive interpretations for decision-makers; and the Monte Carlo
results confirm adequate coverage under repeated sampling. Nevertheless, NSOs should be explicit about the model assumptions underlying the HB estimates, publish direct Bethel-sample estimates as a validation benchmark where feasible, and implement the safeguards described in Sections~\ref{sec:inference} and~\ref{sec:implementation}.

The key message for statistical agencies: by combining Bethel allocation with Hierarchical
Bayes estimation, it is possible to obtain a substantially smaller survey sample --- and
therefore substantially lower operating costs --- while maintaining reliable precision for
all required outputs across all geographic domains.

\section*{Acknowledgements}

The author thanks the developers of the \texttt{R2BEAT} package (Stefano Falorsi, Andrea Fasulo,
Alessio Guandalini, Daniela Pagliuca, Marco D.\ Terribili, Giulio Barcaroli, and Ilaria
Bombelli; primarily affiliated with the Italian National Institute of Statistics, Istat) and the
\texttt{mcmcsae} package (Harm Jan Boonstra; Central Bureau of Statistics, The Netherlands)
for making their software freely available. The author thanks Dr Harm Jan Boonstra and
Dr Andrea Fasulo for helpful advice. The author is grateful to the Associate Editor and one referee for detailed and constructive comments that substantially improved the manuscript.

\bibliographystyle{chicago}

\appendix
\section{Simulation Data-Generating Process}
\label{app:dgp}

The population comprises $N = 1{,}000{,}000$ individuals in $H = 100$ strata nested within
$D = 10$ geographic domains. Stratum design effects $\delta_h \sim U(1.1,1.2)$ are shared
across all three variables.

\subsection{Employment Status}

\[
\eta^E_h = \mathrm{logit}(0.62) + 0.15(X_{1,1,h}-3) + 0.10(X_{1,2,h}-4) + \gamma^E_{d(h)} + \varepsilon^E_h,
\]
where $X_{1,1,h}\sim\mathcal{N}(3,1^2)$, $X_{1,2,h}\sim\mathcal{N}(4,1.5^2)$,
$\gamma^E_d \sim \mathcal{N}(0,0.20^2)$, $\varepsilon^E_h \sim \mathcal{N}(0,0.15^2)$.
Individual indicators $E_{hi}\sim\mathrm{Bernoulli}(\mathrm{logistic}(\eta^E_h))$;
national proportion $\approx 62\%$.

\subsection{Unemployment Status}

\[
\eta^U_h = \mathrm{logit}(0.04) + 0.15(X_{2,1,h}-3) + 0.10(X_{2,2,h}-4) + \gamma^U_{d(h)} + \varepsilon^U_h,
\]
where $X_{2,1,h}\sim\mathcal{N}(3,1^2)$, $X_{2,2,h}\sim\mathcal{N}(4,1.5^2)$ (independent
of employment covariates), $\gamma^U_d \sim \mathcal{N}(0,0.10^2)$,
$\varepsilon^U_h \sim \mathcal{N}(0,0.08^2)$. National proportion $\approx 4\%$.
Mutual exclusivity: $E_{hi} = U_{hi} = 1$ overlaps resolved in ratio $62:4$.

\subsection{Hours Worked}

$\mathrm{Hrs}_{hi}\sim\mathcal{N}(\mu_h, 12^2)$ truncated to $[15,60]$, with
$\mu_h = 15 + 45\,\mathrm{logistic}(0.10 X_{3,1,h} + 0.08 X_{3,2,h})$,
$X_{3,j,h}\sim\mathcal{N}(0,3^2)$. National mean $\approx 37$ hours.

\subsection{Baseline Sample and Bethel Inputs}

A 5\% stratified baseline sample $n^{(0)}_h = \max\{2,\lfloor 0.05 N_h\rceil\}$ provides
stratum means, SDs, and effective sample sizes
$n^{\text{eff}}_h = \lfloor n^{(0)}_h(1-n^{(0)}_h/N_h)/\delta_h \rceil$ as inputs to
\texttt{R2BEAT}.

\subsection{Summary of Parameters}

\begin{table}[ht]
\centering
\caption{Complete summary of simulation parameters (main simulation).}
\label{tab:simparams}
\small
\begin{tabular}{llr}
\toprule
Component & Parameter & Value\\ \midrule
Population     & $N$, $H$, $D$ & $1{,}000{,}000$, 100, 10\\
               & $\delta_h$ & $U(1.1,1.2)$\\ \midrule
Employment     & $\bar{p}_E$; coeff; $\sigma^E_\gamma$; $\sigma^E_\varepsilon$
               & $62\%$; 0.15, 0.10; 0.20; 0.15\\ \midrule
Unemployment   & $\bar{p}_U$; coeff; $\sigma^U_\gamma$; $\sigma^U_\varepsilon$
               & $4\%$; 0.15, 0.10; 0.10; 0.08\\ \midrule
Hours Worked   & $\bar{\mu}$; $\sigma_\varepsilon$; truncation
               & 37 hrs; 12 hrs; $[15,60]$\\ \midrule
Baseline       & fraction; min stratum size & 5\%; 2\\ \midrule
CV targets     & national; domain & 3\%; 8\%\\ \midrule
Monte Carlo    & $B$; iterations (burn-in); $\hat{R}_{\max}$ threshold
               & 1{,}000; 2{,}500 (500); 1.05\\ \bottomrule
\end{tabular}
\end{table}

\section{Sensitivity Scenario Parameters}
\label{app:scenparams}

Table~\ref{tab:scenparams} summarises parameters used in the four sensitivity scenarios
(Section~\ref{sec:sensitivity}). All scenarios use $N = 2{,}000{,}000$, $H = 140$, $D = 13$,
and CV targets of 3\% (national) and 8\% (domain). National prevalences for Scenarios~A--C
were fixed at $E = 65.4\%$, $U = 1.83\%$ by exact numerical calibration of the intercept.

\begin{table}[ht]
\centering
\caption{Sensitivity scenario parameters (new --- added for revision).
$\beta_j$ = covariate coefficient; $\sigma_\gamma$(U) = domain RE SD for unemployment;
$\bar{p}_U$ = national unemployment proportion.}
\label{tab:scenparams}
\small
\begin{tabular}{lcccccc}
\toprule
Scenario & $\beta_1$(E) & $\beta_2$(E) & $\beta_1$(U) & $\beta_2$(U) &
$\sigma_\gamma$(U) & $\bar{p}_U$\\ \midrule
A: Baseline   & 0.150 & 0.100 & 0.150 & 0.100 & 0.10 & 1.83\%\\
B: Weak Aux.  & 0.075 & 0.050 & 0.075 & 0.050 & 0.10 & 1.83\%\\
C: High Het.  & 0.150 & 0.100 & 0.150 & 0.100 & 0.20 & 1.83\%\\
D: Rare Event & 0.150 & 0.100 & 0.150 & 0.100 & 0.10 & 0.50\%\\ \bottomrule
\end{tabular}
\end{table}

\end{document}